\documentclass[journal, twoside, 10pt]{IEEEtran}

\ifCLASSOPTIONcompsoc
\usepackage[nocompress]{cite}
\else
\usepackage{cite}
\fi
\usepackage{color}
\usepackage{multirow,threeparttable}
\usepackage{graphicx}
\usepackage{amsthm,amsmath,amssymb}
\usepackage{mathrsfs}
\usepackage{bm}
\usepackage{algorithm}
\usepackage{algorithmic}

\usepackage{booktabs}
\usepackage{array}
\usepackage{enumitem}
\usepackage{subfigure}
\usepackage{multirow}
\usepackage{gensymb}

\usepackage{amsmath}
\begin{document}

\title{A Point-Hyperplane Geometry Method for Operational Security Region of Renewable Energy Generation in Power Systems}
%
\author{Can~Wan,~\IEEEmembership{Senior Member,~IEEE,}~
        Biao~Li,~
        Xuejun~Hu,~
        Yunyi~Li,~
        Ping~Ju,~\IEEEmembership{Senior Member,~IEEE}
\thanks{
This work was support by National Natural Science Foundation of China under Grants U2066601.   

C. Wan, B. Li, X. Hu, Y. Li, and P. Ju are with the College of Electrical Engineering, Zhejiang University, Hangzhou 310027, China (e-mails: canwan@zju.edu.cn; biaoli@zju.edu.cn; huxuejun@zju.edu.cn; yunyi\_li@zju.edu.cn, pju@hhu.edu.cn). 
}

}

\maketitle

\markboth{IEEE TRANSACTIONS ON POWER SYSTEMS}
{Can \MakeLowercase{\textit{et al.}}: A Point-Hyperplane Geometry Method for Operational Security Region of Renewable Energy Generation}

\begin{abstract}
  The rapid growth of renewable energy generation challenges the secure operation of power systems. 
  It becomes crucial to quantify the critical security boundaries and hosting capability of renewable generation at the system operation level. 
  This paper proposes a novel point-hyperplane geometry (PHG) method to accurately obtain the geometric expression of the operational security region of renewable energy generation for power systems.
  Firstly, the geometric expression of the operational security region is defined as a polytope of boundary hyperplanes in the form of inequalities satisfying the system operation constraints. 
  Then, an orthogonal basis generation method is proposed to solve a single boundary hyperplane of the polytope based on intersecting and orthogonal geometric principles. 
  Next, a point-hyperplane iteration algorithm is developed to progressively obtain the overall geometric polytope of the operational security region of renewable energy generation in power systems.
  Besides, the flexible performance trade-off can be achieved by modifying the proposed maximum tolerated angle between adjacent hyperplanes.
  Finally, comprehensive case stydies verify the effectiveness and superiority of the PHG method.

\end{abstract}

\IEEEpeerreviewmaketitle

\section{Introduction}
\IEEEPARstart{R}{enewable} energy generation (REG) has been playing a vital role in mitigating the energy crisis. However, the rapid growth of REG with high uncertainty threatens the secure operation of power systems \cite{8635327}. It is crucial to describe the security boundaries of REG outputs in the power systems quantifiably.
Thus, the concept of security region has been proposed in various scenarios of power systems and also provided important decision or assessment support for the control, dispatching, planning, etc\cite{7353226, 10418517}. 

The security region can be divided into two categories at the physical process level: steady-state security region \cite{1085091,6547732,8454506} and dynamic security region \cite{WuHao2021,he_dynamic_2024,9280366}. Considering the influence of the components in power systems, there are also classifications such as N-0 security region (under normal operating conditions) \cite{8454506,wei_dispatchable_2015}, N-1 security region \cite{6570751,8723179,9540287}, and so on. With the development of the REG on the demand side, some studies on security regions have also recently been extended from the transmission level to the distribution level \cite{8026182, wen_aggregate_2022 ,8314113}. 
This study mainly focuses on the N-0 operational security region of REG outputs under the steady-state transmission-level operation, which can be used to quantify the inherent hosting capacity and accommodation level of REG in power systems.

Essentially, the security region is the subregion space composed of sub-variables (active outputs, voltage, etc) under the feasible region of the power flow constraints in the power system. It is widely applicable to various research objects and has different expressions \cite{wei_dispatchable_2015,wang_sequential_2021}.
For examples, the hyperplane and vertices double description is introduced to model and quantify the maximum uncertainty boundary and operation region of distributed generation uncertainties \cite{wan_maximum_2018}.
A robust security region is built based on convex hull form to obtain a set of uniformly distributed points on nonlinear power flow solvability boundaries and creates additional dimensions of flexibility of system operation\cite{8581491}. 
In general, geometric expressions for nonlinear operational security regions are implicit and difficult to obtain.

Therefore, the linearized ones are explicit and concise alternatives. For solving them, The Fourier-Motzkin elimination method is a traditional solution \cite{7442895} that systematically eliminates variables from a system of linear inequalities to produce a reduced set of inequalities in the remaining variables. However, its computation time increases exponentially as the system scale grows, and there are several methods to speed up like Chernikove rules \cite{FMEA1} and bit pattern trees \cite{FMEA2}. 
Also, multi-parametric programming methods have been proposed, which construct multi-parametric combinations and regional linear mapping \cite{DAI2019115975}. For example, a multi-parametric programming-based fast boundary search model is proposed to validate the power transmission region \cite{8889678}. However, there is still a problem of low computational efficiency due to too many constraint combinations when the system is complex.

This paper proposes a novel point-hyperplane geometry (PHG) method to accurately obtain the geometric expression of the operational security region of REG, which provides a general geometric solution that differs from the traditional methods. The operational security region is modeled as a geometry constructed from boundary hyperplanes, which is efficiently based on the adjacency geometric relationship. 
First, the target geometric operational security region is defined as a polytope.  
Then, an orthogonal basis generation method is proposed to solve a boundary hyperplane of the polytope based on a given exterior point, based on intersecting and orthogonal geometric principles.
Next, a point-hyperplane iteration algorithm is developed to search all boundary hyperplanes, which has a basic iterative structure of exterior point-adjacent boundary hyperplane-overall geometry. A dynamic exterior point adjustment method is proposed to ensure iterative continuity by avoiding falling into unsolvable scenarios.  The efficiency of iteration is enhanced by a progressive geometric expansion method. Besides, a maximum tolerated angle is presented to make a balance between accuracy and computational cost.
Finally, some case studies demonstrate the accuracy, sensitivity, and efficiency of the proposed PHG method.

The major contributions of this paper are summarized below: 
\vspace{-0.1em}
\begin{enumerate}[leftmargin=*]
  \item
  A PHG method is proposed to obtain the geometric expression of the operational security region of renewable energy generation, which uses a series of boundary hyperplanes to progressively construct the geometry.
 \item
  An orthogonal basis generation method is proposed to solve a single boundary hyperplane of the geometry, which is formulated as linear programming based on intersecting and orthogonal geometric principles.
 \item
  A point-hyperplane iteration algorithm is developed to effectively calculate all boundary hyperplanes of the operational security region by dynamically adjusting unsolvable points and expanding adjacent hyperplanes. 
 \item
  A maximum tolerated angle between adjacent boundary hyperplanes is presented to provide a flexible performance tradeoff in accuracy and computational costs.
\end{enumerate}
\vspace{-0.1em}

\section{Operational Security Region of Renewable Energy Generation: Definition}
\subsection{Feasible Region of AC Power Systems}
The feasible region of an AC power system is non-linear and non-convex, which is denoted as $\mathbb{F}^{\rm AC}$ and determined by nodal power flow equations \eqref{nodal power flow equations1}-\eqref{nodal power flow equations2}, upper and lower limits of power generation output \eqref{generation output limitation1}, unit ramp rate constraints \eqref{unit ramp rate constraints1}, voltage limitations \eqref{voltage limitations} and branch flow constraints \eqref{branch flow constraints}, expressed as
 
\begin{IEEEeqnarray}{l}
  \label{nodal power flow equations1}
  \hspace{-0.5em} p_{i}=\sum_{j \in \mathcal{N}} v_{i} v_{j}\left(G_{i j} \cos \theta_{i j}+B_{i j} \sin \theta_{i j}\right), \ \forall i\in\mathcal{N}\\
  \label{nodal power flow equations2}
  \hspace{-0.5em} q_{i}=\sum_{j \in \mathcal{N}} v_{i} v_{j}\left(G_{i j} \sin \theta_{i j}-B_{i j} \cos \theta_{i j}\right),\ \forall i\in\mathcal{N}\\
  \label{generation output limitation1}
  \hspace{-0.5em} \underline{p}_{i}^{\rm g} \leq p_{i}^{\rm g} \leq \overline{p}_{i}^{\rm g},\ 
  \underline{q}_{i}^{\rm g} \leq q_{i}^{\rm g} \leq \overline{q}_{i}^{\rm g},\ \forall i\in\mathcal{G}\\
  \label{unit ramp rate constraints1}
  \hspace{-0.5em} p_{i}^{\rm g}-p_{i}^{\rm g, last} \leq RU_{i}^{\rm g},\ p_{i}^{\rm g}-p_{i}^{\rm g, last} \geq RD_{i}^{\rm g},\ \forall i\in\mathcal{G}\\
  \label{voltage limitations}
  \hspace{-0.5em} \underline{v}_{i} \leq v_{i} \leq \overline{v}_{i},\ \forall i\in\mathcal{N}\\
  \label{branch flow constraints}
  \hspace{-0.5em} \underline{p}_{ij} \leq p_{ij} \leq \overline{p}_{ij},\ \forall (i,j)\in\mathcal{L}
\end{IEEEeqnarray}
where the overline and underline denote the upper and lower limits of variables; $\mathcal{N}$ is the set of nodes in the power system and $\mathcal{G}$ is the set of generator nodes in the power system; $\mathcal{L}$ denotes the set of branch lines; $v_i$ and $\theta_i$ are the voltage magnitude and phase angle at node $i$, respectively; $\theta_{i j}=\theta_{i}-\theta_{j}$ refers to the phase angle difference between node $i$ and node $j$; $G_{i j}$ and $B_{i j}$ are the conductance and susceptance between node $i$ and node $j$, respectively; $p_{i}+jq_{i}$ is the complex injection power at node $i$; $p_{i}^{\rm g}+jq_{i}^{\rm g}$ refers to the complex generation power at node $i$; $RU_{i}^{\rm g}$ and $RD_{i}^{\rm g}$ are the up and down ramping limit of generator $i$, respectively; $p_{i}^{\rm g, last}$ is the active generation power at node $i$ at last operation time; $p_{ij}$ and $q_{ij}$ represent the active and reactive branch flow from node $i$ to node $j$, respectively, which are given by
\begin{IEEEeqnarray}{l}
  \label{branch flow equation}
  p_{i j}=v_{i} v_{j}\left(G_{i j} \cos \theta_{i j}+B_{i j} \sin \theta_{i j}\right)-G_{i j} v_{i}^{2}\\
  q_{i j}=v_{i} v_{j}\left(G_{i j} \sin \theta_{i j}-B_{i j} \cos \theta_{i j}\right)+B_{i j} v_{i}^{2}
\end{IEEEeqnarray}

In addition, considering a node connected with generators, REGs and loads, the injection power can be formulated as
\begin{IEEEeqnarray}{l}
  p_{i}=p_{i}^{\rm g}+w_{i}-p_{i}^{\rm d}
\end{IEEEeqnarray}
where $w_{i}$ and $p_{i}^{\rm d}$ are the active power output of renewable energy and active power demand at node $i$, respectively. 

To analyze the feasible region for power systems, the outputs of REG are regarded as random variables. The outputs of REG satisfy:
\begin{IEEEeqnarray}{l}
  \label{output constraints}
  w_{i}\geq 0, \ \forall i\in\mathcal{R}
\end{IEEEeqnarray}
where $\mathcal{R}$ is the set of nodes equipped with REG.

Then the feasible region $\mathbb{F}^{\rm AC}$ is given by:
\begin{align}
  \mathbb{F}^{\rm AC}:=\{(w,x)|\
  f(w,x)=0,\ \ g(w,x)\leq 0\}
\end{align}
where $w$ represents the output of REG, i.e. $w:=(w_{i},\ i\in\mathcal{R})$; $x$ is other varibles of the power system, i.e. $x:=(p_{i}^{\rm g}, q_{i}^{\rm g}, \ i\in\mathcal{G}; v_i, \theta_i, \ i\in\mathcal{N})$; $f(w,x)=0$ refer to the nodal power balance equations; $g(w,x)\leq 0$ represent the inequality constraints.

Based on the feasible region $\mathbb{F}^{\rm AC}$, an ideal operational security region (IOSR) $\mathbb{S}^{\rm AC}$ of REG is defined, consisting only of REG outputs $w$. The $\mathbb{S}^{\rm AC}$ for AC power systems has the following two definitions: 

1) $\mathbb{S}^{\rm AC}$ is a region represented only by $w$, given by
\begin{align}
\mathbb{S}^{\rm AC}:=\{w|\ h(w)\leq 0\}
\end{align}

2) For $w$, the relationship between  $\mathbb{S}^{\rm AC}$ and $\mathbb{F}^{\rm AC}$ is expressed as
\begin{IEEEeqnarray}{l}
  w \in \mathbb{S}^{\rm AC} \Leftrightarrow \exists x \rightarrow (w, x) \in \mathbb{F}^{\rm AC}
 \end{IEEEeqnarray}
where $h(w)\leq 0$ refers to all equality and inequality constraints about $w$ and it only has an implicit form. 
Mathematically, $\mathbb{S}^{\rm AC}$ is essentially the projection of the $\mathbb{F}^{\rm AC}$ onto the sub-space of REG output $w$. However, $\mathbb{F}^{\rm AC}$ is a non-convex and non-linear region, and the expression of its projection $\mathbb{S}^{\rm AC}$ is extremely complex and difficult to solve.

\subsection{Geometric Operational Security Region of REG}
Using linearization methods, the non-linear and non-convex constraints of the AC power system in \eqref{nodal power flow equations1}, \eqref{nodal power flow equations2}, and \eqref{branch flow constraints} can be converted into linear ones. The first-order Taylor series approximation method is utilized to linearize the AC power flow constraints under a given Taylor expansion point $(v^{0},\theta^{0})=(1,0)$. Then the non-linear and non-convex constraints \eqref{nodal power flow equations1}, \eqref{nodal power flow equations2}, and \eqref{branch flow constraints} turn into:
\begin{IEEEeqnarray}{l}
  \label{Linearized nodal power1}
  p_{i}=\sum_{j \in \mathcal{N}} G_{i j}\left(v_{i}+v_{j}-1\right)+B_{i j} \theta_{i j},\ \forall i\in\mathcal{N} \\ 
  \label{Linearized nodal power2}
  q_{i}=\sum_{i \in \mathcal{N}} G_{i j} \theta_{i j}-B_{i j}\left(v_{i}+v_{j}-1\right),\ \forall i\in\mathcal{N} \\
  \label{Linearized branch flow}
  \underline{p}_{ij} \leq G_{i j}\left(v_{j}-v_{i}\right)+B_{i j} \theta_{i j} \leq \overline{p}_{ij},\ \forall (i,j)\in\mathcal{L}
\end{IEEEeqnarray}

Then the linearized feasible region $\mathbb{F}^{\rm L}$ is given by
\begin{align}
  \label{FL}
  \mathbb{F}^{\rm L}:=\{(w,x)|\
  &A_{i}w+B_{1}x+F_{1}=0,\notag\\
  &A_{2}w+B_{2}x+F_{2}\leq 0\}
\end{align}
where $A_{1}w+B_{1}x+F_{1}=0$ refers to the the equality constraints and $A_{2}w+B_{2}x+F_{2}\leq 0$ indicates the inequality constraints of the linearized AC power system. The linearized feasible region $\mathbb{F}^{\rm L}$ is a convex polytope in a linear space of all variables.

As the projection of the linearized feasible region $\mathbb{F}^{\rm L}$ onto the space of REG, a geometric operational security region (GOSR) is proposed.  Since the projection of a convex region is still a convex space, the GOSR $\mathbb{S}^{\rm L}$ is also a convex polytope, which consists of linear constraints, given by
\begin{align}
\mathbb{S}^{\rm L}:=\{w|\ Cw+D\leq 0\}
\end{align}
where the GOSR $\mathbb{S}^{\rm L}$ also satisfies 
\begin{IEEEeqnarray}{l}
  \label{relation of V-AC and S-AC}
  w \in \mathbb{S}^{\rm L} \Leftrightarrow \exists x \rightarrow (w, x) \in \mathbb{F}^{\rm L}
\end{IEEEeqnarray}
and it is evident that the $\mathbb{S}^{\rm L}$ has a explicit and concise geometric expression and more convenient to be dealt with than the non-convex and non-linear IOSR $\mathbb{S}^{\rm AC}$. 
Therefore, the GOSR $\mathbb{S}^{\rm L}$ is the target polytope expected to be solved by the  proposed PHG method in this paper.

\section{Orthogonal Basis Generation Method for a Single Boundary Hyperplane}

\subsection{Boundary Point Solving}

 Each boundary hyperplane of the GOSR corresponds to an inequality of the GOSR $\mathbb{S}^{\rm L}$. Suppose the GOSR $\mathbb{S}^{\rm L}$ is consists of K inequalities, i.e. $ \mathbb{S}^{\rm L}:=\{w|c_i^{T} w+d_i\leq 0, i=1,2,\dots,K\}$. Then the K boundary hyperplanes are given by
\begin{IEEEeqnarray}{l}
  H_i: c_i^{T} w+d_i= 0, i=1,2,\dots,K
\end{IEEEeqnarray}
where $c_i^{T}$ is the $i$-row vector of $C$ and $d_i$ is the $i$-th element of $D$ in inequalities of the GOSR $\mathbb{S}^{\rm L}$.
Before solving a boundary hyperplane $H_i$ of the GOSR $\mathbb{S}^{\rm L}$, it is necessary to obtain a point on it, denoted as a boundary point. 

A boundary point solving (BPS) model is proposed to search a boundary point of a boundary hyperplane that intersects the line segment between an arbitrary given point $w^{\rm gp}$ and a given interior point $w^{\rm in}$, which serves to 1) determining whether the $w^{\rm gp}$ is an interior point ($w \in \mathbb{S}^{\rm L}$) or an exterior point ($w \notin \mathbb{S}^{\rm L}$); 2) generating the boundary point if $w^{\rm gp}$ is an exterior point. The BPS model is formulated as linear programming that is constrained by power system operation security, given by
\begin{IEEEeqnarray}{l}
  \label{BPS model}
  \IEEEyesnumber\IEEEyessubnumber*
  \min_{\lambda, x}\ -\lambda \\
  \label{BPS constraints}
  s.t.\ (w^{\rm in}+(w^{\rm gp}-w^{\rm in})\lambda,x)\in\mathbb{F}^{\rm L},\ 0\leq \lambda\leq 1
\end{IEEEeqnarray}
where $w^{\rm in}$ is selected as the coordinate origin in this paper and the point on the line segment between the given point $w^{\rm gp}$ and the interior point $w^{\rm in}$ is formulated by $w^{\rm in}+(w^{\rm gp}-w^{\rm in})\lambda,\  0\leq \lambda\leq 1$, where $\lambda$ is the scaling parameter \cite{lang2019introduction} and \eqref{BPS constraints} refers to \eqref{FL}.

Since the BPS model in \eqref{BPS model} is convex programming, its solution $(\lambda^{*}, x^{*})$ is a global optimum. In addition, for any convex region, all points on a line segment connecting two interior points are interior points. Then the given point $w^{\rm gp}$ belongs to the GOSR $\mathbb{S}^{\rm L}$ if and only if $\lambda^{*}=1$. Therefore, the BPS model serves to determine whether the given point $w^{\rm gp}$ is an interior point or an exterior point of the GOSR $\mathbb{S}^{\rm L}$, shown as following
\begin{IEEEeqnarray}{l}
  \begin{cases}
    w^{\rm gp} (interior)  \quad \lambda^{*}=1\\
    w^{\rm gp} (exterior)  \quad \lambda^{*}<1
  \end{cases}
\end{IEEEeqnarray}

If the given point $w^{\rm gp}$ is an exterior point, there exists a unique boundary point of the GOSR $\mathbb{S}^{\rm L}$, which lies on the line segment between the given point $w^{\rm gp}$ and the coordinate origin $w^{\rm in}$ and is not any endpoint of this line segment. This boundary point can also be generated by the BPS model. In fact, the solved point $w^{\rm b}=w^{\rm in}+(w^{\rm gp}-w^{\rm in})\lambda^{*}$ is a boundary point of the GOSR $\mathbb{S}^{\rm L}$, which described as \textit{Proposition 1}.

\textit{Proposition 1}: Suppose the given point $w^{\rm gp}$ is an exterior point of the GOSR $\mathbb{S}^{\rm L}$, and there exists a necessary and sufficient condition in \eqref{relation of V-AC and S-AC}, then the point $w^{\rm b}$ solved by \eqref{BPS model} is a point on a boundary hyperplane of $\mathbb{S}^{\rm L}$.

\textit{Proof}: The proof of \textit{Proposition 1} is given in Appendix \ref{Proof of Proposition 1}.

\subsection{Orthogonal Basis Generation}
Based on the specific boundary point obtained by the BPS model, an orthogonal basis generation (OBG) method is further developed to solve the corresponding boundary hyperplane in $n-$dimensional linear space consisting of $n$ REG nodes in the power system, by generating orthogonal basis vectors and obtaining other $n-1$ new boundary points. The further OBG method consists of three major steps:

\subsubsection{Uniqueness and Correctness of Solved Boundary Hyperplane}
Suppose that the dimension of REG optputs $w$ is $n$, i.e. $n=|\mathcal{R}|$. \textit{Theorem 1} gives the necessary and sufficient condition for using $n$ boundary points to determine a corresponding boundary hyperplane uniquely. 

\textit{Theorem 1}: The necessary and sufficient condition for using $n$ points ($\alpha_{n},\ i=1,2,\dots,n$) to uniquely determine a hyperplane in the $n$-dimensional linear space is that $n-1$ basis vectors $\alpha_{2}-\alpha_{1},\alpha_{3}-\alpha_{1},\dots,\alpha_{n}-\alpha_{1}$ are linearly independent ($n-1$ orthogonal basis).

\textit{Proof}: The proof is given in Appendix \ref{Proof of Theorem 1}.

Therefore, to ensure the uniqueness and correctness of the boundary hyperplane, the $n$ boundary points (a boundary point obtained by BPS model and other $n-1$ boundary points solved by the further OBG method) are necessary for uniquely determining a boundary hyperplane.

\begin{figure}[tb]
  \centering
  \includegraphics[width=3.3in]{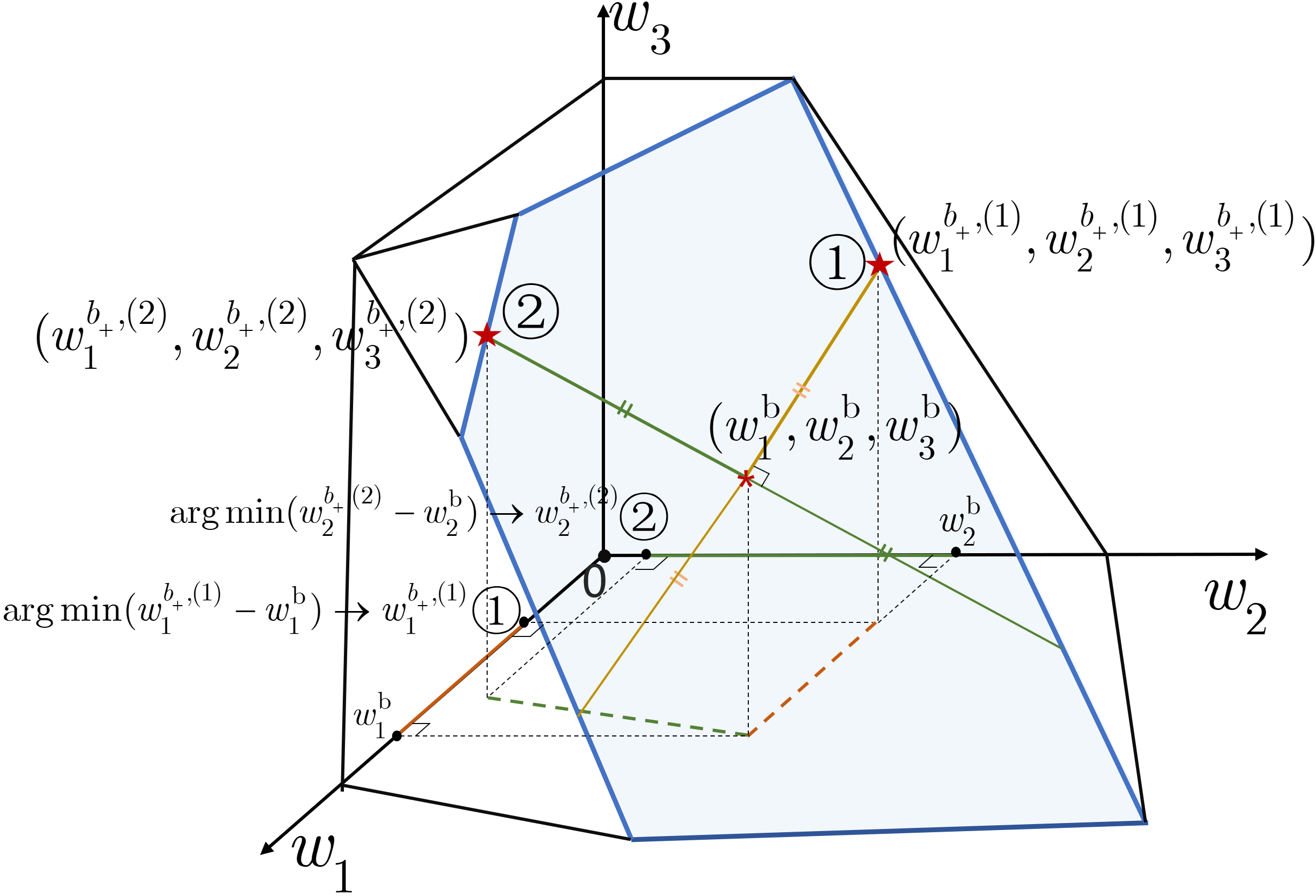}
  \caption{Schematic diagram of proposed OBG method}
  \label{Fig.3D_diagram}
\end{figure}

\subsubsection{Orthogonal Basis Vectors Generation}
Denote $w^{{\rm b}_+, (i)},\ i=1,2,\dots,n-1$ as the $n-1$ new boundary points. According to \textit{Theorem 1}, to uniquely determine a boundary hyperplane passing through the boundary point $w^{\rm b}$, these new $n-1$ boundary points are required to satisfy: (1) \textit{C1}: they are on the boundary hyperplane of the GOSR $\mathbb{S}^{\rm L}$ passing through $w^{\rm b}$; (2) \textit{C2}: the vectors $w^{{\rm b}_+, (i)}-w^{\rm b}, \ i=1,2,\dots,n-1$ are linearly independent.

\textit{Proposition 2}: Suppose that $w^{\rm b}$ is a boundary point of the GOSR $\mathbb{S}^{\rm L}$ and the necessary and sufficient condition in \eqref{relation of V-AC and S-AC} holds. If there exists a point $w^{{\rm b}_+}$ satisfying $(w^{{\rm b}_+}, x^{+})\in\mathbb{F}^{\rm L}$, $(w^{{\rm b}_-}, x^{-})\in\mathbb{F}^{\rm L}$, and $w^{{\rm b}_-}=2 w^{\rm b}-w^{{\rm b}_+}$, where $w^{{\rm b}_-}, x^{-}$ is the auxiliary point, then the new point $w^{{\rm b}_+}$ is also on the hyperplane of $\mathbb{S}^{\rm L}$ passing through the boundary point $w^{\rm b}$.

\textit{Proof}: The proof of \textit{Proposition 2} is given in Appendix \ref{Proof of Proposition 2}.

From \textit{Proposition 2}, it is shown that the points $w^{{\rm b}_+, (i)},\ i=1,2,\dots,n-1$ are ensured to satisfy condition \textit{C1} by establishing following constraints for all new boundary points:
\begin{IEEEeqnarray}{l}
  \label{constraints-on hyperplane}
  (w^{{\rm b}_-,(i)}, x^{-})\in\mathbb{F}^{\rm L},\ 
  w^{{\rm b}_-,(i)}=2 w^{\rm b}-w^{{\rm b}_+,(i)},\notag\\
  (w^{{\rm b}_+,(i)}, x^{+})\in\mathbb{F}^{\rm L},\ \forall i\in\{1,2,\dots,n-1\}
\end{IEEEeqnarray}

Mathematically, orthogonal vectors are linearly independent \cite{axler2024linear}. Therefore, the vectors $w^{{\rm b}_+, (i)}-w^{\rm b}, \ i=1,2,\dots,n-1$ can be designed into orthogonal basis vectors to guarantee the new boundary points satisfying condition \textit{C2}, given by 
\begin{IEEEeqnarray}{l}
  \label{orthogonal conditions-nonconvex}
  (w^{{\rm b}_+, (j)}-w^{\rm b})^{T}(w^{{\rm b}_+, (i)}-w^{\rm b})=0,\notag\\
  \quad\quad\quad\quad\quad\quad\quad\quad \forall \ i,j \in \{1,2,\dots,n-1\},\ i\neq j
\end{IEEEeqnarray}

The orthogonal conditions \eqref{orthogonal conditions-nonconvex} are non-linear equations. Since these new boundary points are obtained successively in the OBG method, the information of the previous points is known for the $i$-th point. Therefore, the orthogonal conditions \eqref{orthogonal conditions-nonconvex} can be equivalently converted into following linear equations for the $i$-th ($i\geq2$) new boundary point $w^{{\rm b}_+,(i)}$:
\begin{IEEEeqnarray}{l}
  \label{orthogonal conditions-linear-each}
  (w^{{\rm b}_+,(j)}-w^{\rm b})^{T} (w^{{\rm b}_+,(i)}-w^{\rm b})=0,\notag\\
  \quad\quad \quad\quad \quad\quad\quad\quad \forall j\in \{1,2,\dots,i-1\},\ i\geq 2 
\end{IEEEeqnarray}

If $n-1$ new boundary points are generated, the vectors $w^{{\rm b}_+, (i)}-w^{\rm b}, \ i=1,2,\dots,n-1$ satisfy 
\begin{IEEEeqnarray}{l}
  \label{orthogonal conditions-linear}
  (w^{{\rm b}_+,(j)}-w^{\rm b})^{T} (w^{{\rm b}_+,(i)}-w^{\rm b})=0,\notag\\
  \quad\quad\quad\ \forall j\in \{1,2,\dots,i-1\},\ i\in\{2,3,\dots,n-1\}
\end{IEEEeqnarray}

It is indicated from \eqref{orthogonal conditions-linear} that the vectors are pairwise orthogonal, which is equivalent to \eqref{orthogonal conditions-nonconvex}. Therefore, the condition \textit{C2} are satisfied by using the linear constraints in \eqref{orthogonal conditions-linear-each} for solving the $i$-th new boundary point $w^{{\rm b}_+,(i)}$. 
Certainly, the boundary hyperplane passing through the boundary point $w^{{\rm b}}$ is uniquely determined by these $n-1$ non-zero vectors.

Then, a new boundary point generation (NBPG) model is proposed to obtain each new boundary point (suppose it is the $i$-th point $w^{{\rm b}_+, (i)}$), the main purpose of which is to guarantee that the vector $w^{{\rm b}_+, (i)}-w^{\rm b}$ is non-zero. The NBPG model to solve $i$-th new boundary point is formulated as linear programming in \eqref{boundary hyperplane Solving model} based on operational constraints, which ensures that the solution for the $i$-th new boundary point is global optimal. 
\begin{IEEEeqnarray}{l}
  \label{boundary hyperplane Solving model}
  \IEEEyesnumber\IEEEyessubnumber*
  \min_{w^{{\rm b}_+,(i)},x^{+},x^{-}}\ w_{i}^{{\rm b}_+,(i)}-w_{i}^{\rm b}\\
  \label{orthogonal constraints}
  s.t.\ (w^{{\rm b}_+,(j)}-w^{\rm b})^{T} (w^{{\rm b}_+,(i)}-w^{\rm b})=0,\notag\\
  \quad\quad\quad\quad\quad\quad \forall j\in \{1,2,\dots,i-1\},\ i\geq 2\\
  \label{Constraint1:points on hyperplane}
  \quad\quad (w^{{\rm b}_+,(i)}, x^{+})\in\mathbb{F}^{\rm L},\
  (w^{{\rm b}_-,(i)}, x^{-})\in\mathbb{F}^{\rm L}\\
  \label{Constraint3:points on hyperplane}
  \quad\quad w^{{\rm b}_-,(i)}=2 w^{\rm b}-w^{{\rm b}_+,(i)}
\end{IEEEeqnarray}
where $w_{i}^{{\rm b}_+,(i)}$ and $w_{i}^{\rm b}$ are the coordinates of the points $w^{{\rm b}_+,(i)}$ and $w^{\rm b}$ on the $i$-th coordinate axis, respectively, and the constraint \eqref{Constraint1:points on hyperplane} refers to \eqref{FL}.

\textcolor{blue}{Fig. \ref{Fig.3D_diagram}} depicts the determination process of each new boundary point on a certain boundary hyperplane (the “blue” plane) in a 3-dimensional linear space. The point $w^{\rm b}$ (marked by ``$*$") is a given boundary point. Obviously, the first new boundary point $w^{{\rm b}_+,(1)}$ are generated without orthogonal constraints in \eqref{orthogonal constraints} due to $i=1$, and it only needs to satisfy: 1) the coordinate of $w^{{\rm b}_+,(1)}$ on the axis $w_{1}$ is the smallest axis on the ``blue" plane; 2) there is also a point $w^{{\rm b}_-,(1)}$ on the ``blue" plane such that $w^{\rm b}$ is their midpoint. Then, in addition to satisfying these two conditions, the second new boundary point $w^{{\rm b}_+,(2)}$ also needs to satisfy an orthogonal condition, i.e. it lies on the vertical bisector of the line segment between the new boundary point $w^{{\rm b}_+,(1)}$ and auxiliary point $w^{{\rm b}_-,(1)}$. Then the two points are successfully obtained, and the three points $w^{\rm b}, w^{{\rm b}_+,(1)},\ w^{{\rm b}_+,(2)}$ satisfy the necessary and sufficient condition in \textit{Theorem 1} to uniquely determine a hyperplane in a 3-dimensional linear space of REG.

\subsubsection{Geometric Expression of Boundary Hyperplane}

A hyperplane in an $n$-dimensional linear space can be typically formulated as $H: c^T w+d=0$, where the dimensions of $c$ and $d$ are $n$ and 1, respectively. After obtaining $n-1$ new boundary points $w^{{\rm b}_+,(i)}, \ i=1,2,\dots,n-1$, the boundary hyperplane passing through the boundary point $w^{\rm b}$ is determined by solving a homogeneous system of linear equations, given by
\begin{align}
  \label{HD hyperplane}
  \left[
  \begin{array}{ll}
    (w^{\rm b})^{T} &1 \\
    (w^{{\rm b}_+,(1)})^{T} &1\\
    (w^{{\rm b}_+,(2)})^{T} &1\\
    \dots & \dots\\
    (w^{{\rm b}_+,(n-1)})^{T} &1
  \end{array}
  \right]
  \left[
  \begin{array}{l}
    c\\
    d
  \end{array}
  \right]
  =0
\end{align}
where the non-zero solutions of $[c,d]$ determine the hyperplane expression $H: c^{T} w+d= 0$.

\textit{Lemma 1}: Suppose points $w^{{\rm b}_+,(i)},\ i=1,\dots,n-1$ in an $n$-dimensional linear space satisfy:

\textit{L1}: The $i$-th new boundary point $w^{{\rm b}_+,(i)}$ is solved by \eqref{boundary hyperplane Solving model};

\textit{L2}: $\|w^{{\rm b}_+,(i)}-w^{\rm b}\| > 0,\ \forall i=1,\dots,n-1$.

Then the hyperplane that uniquely solved by \eqref{HD hyperplane} is a single boundary hyperplane of the GOSR $\mathbb{S}^{\rm L}$ passing through the boundary point $w^{\rm b}$.

\textit{Proof}: The proof of \textit{Lemma 1} is given in Appendix \ref{Proof of Lemma 1}.

In fact, the condition \textit{L2} is used to ensure the vectors $w^{{\rm b}_+, (i)}-w^{\rm b}, \ i=1,2,\dots,n-1$ are non-zero, which also means that these $n-1$ points are different points. Then \textit{Lemma 1} shows that the boundary hyperplane is uniquely solved by \eqref{HD hyperplane} if the points is obtained by the NBPG model and their objective functions are all non-zero.

\section{Point-Hyperplane Iteration Algorithm for GOSR}

\subsection{Dynamic Exterior Point Adjustment}
Based on the OBG method, a point-hyperplane iteration (PHI) algorithm is proposed to obtain all the boundary hyperplanes of the GOSR. 
From a given interior point and the exterior point, a boundary point can be obtained by the BPS model, and the corresponding boundary hyperplane is solved by the rest parts of the OBG method. As mentioned in \textit{Lemma 1}, the condition \textit{L2} is necessary for uniquely obtaining a boundary hyperplane. However, the condition \textit{L2} may not be satisfied in some cases. Under a given boundary point $w^{\rm b}$, if the condition \textit{L2} are not satisfied by new boundary points solved by the OBG method, it is a ``bad" boundary point $w^{\rm b,bad}$ and the corresponding exterior point is a ``bad" exterior point $w^{\rm ex,bad}$. 

\begin{figure}[tb]
  \centering
  \includegraphics[width=3.6in]{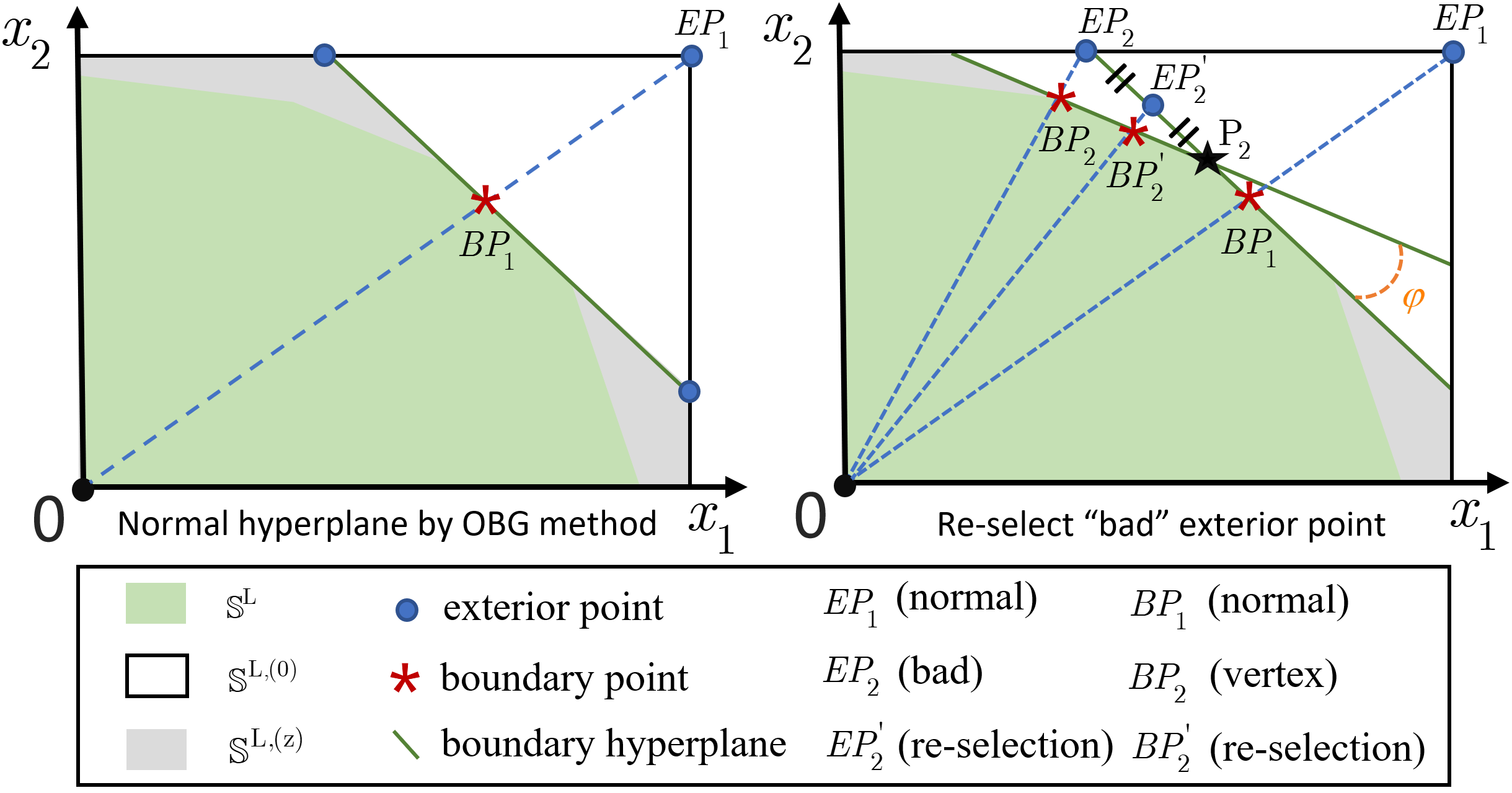}
  \caption{Schematic diagram of DEPA method}
  \label{Fig.re-selection}
\end{figure}

Therefore, a dynamic exterior point adjustment (DEPA) method is established to ensure iterative continuity and reliability of the incremental boundary hyperplane by obtaining a suitable re-selection exterior point $\overline{w}^{\rm ex}$.
\textcolor{blue}{Fig. \ref{Fig.re-selection}} gives an example of a ``bad" exterior point situation, and describes the DEPA method for dynamically adjusting $\overline{w}^{\rm ex}$. The point ${\rm EP}_{2}$ is a $w^{\rm ex,bad}$, because the boundary point ${\rm BP}_{2}$ on the line segment ${\rm EP}_{2}$ to ${\rm O}$ is a vertex. Based on a vertex as the boundary point, the $i$-th new boundary point $w^{{\rm b}_+,(i)}$ solved by the NBPG model must violate condition \textit{L2}, thereby it is unable to generate a boundary hyperplane.

Therefore, the DEPA method firstly solves such a boundary point (${\rm P}_{2}$), which satisfies: (1) it belongs to the SOGR $\mathbb{S}^{\rm L}$; (2) it is on the line segment between the exterior point ${\rm EP}_{2}$ and the previous boundary point ${\rm BP}_{1}$; (3) it is farthest from the previous boundary point ${\rm BP}_{1}$. 
Here, ${\rm BP}_{1}$ is the previous boundary point $w^{\rm b, pre}$ and ${\rm EP}_{2}$ is the ``bad" exterior point $w^{\rm ex, bad}$. The point on the line segment between these two points are formulated as $w^{\rm b, pre}+\lambda^{\rm p} (w^{\rm b, bad}-w^{\rm b, pre}),\ 0\leq\lambda^{\rm p}\leq 1$, where $\lambda^{\rm p}$ is the scaling parameter. Then the boundary point $w^{\rm b, p *}$ (${\rm P}_{2}$) is the optimal solution to \eqref{farthest point}, given by $w^{\rm b, p *}=w^{\rm b, pre}+\lambda^{{\rm p}*} (w^{\rm b, bad}-w^{\rm b, pre})$.
\begin{IEEEeqnarray}{l}
  \label{farthest point}
  \IEEEyesnumber\IEEEyessubnumber*
  \max_{\lambda^{\rm p},x}\ \lambda^{\rm p},\quad  0\leq\lambda^{\rm p}\leq 1\\
  s.t.\  (w^{\rm b, p}, x)\in\mathbb{F}^{\rm L}\\
  \quad\quad w^{\rm b, p}=w^{\rm b, pre}+\lambda^{\rm p} (w^{\rm b, bad}-w^{\rm b, pre})
\end{IEEEeqnarray}

Next, the re-selection exterior point $\overline{w}^{\rm ex}$ is adjusted to the midpoint between $w^{\rm ex, bad}$ and $w^{\rm b, p *}$, formulated as $\overline{w}^{\rm ex}=\frac{1}{2}(w^{\rm ex, bad}+w^{\rm b, p *})$. Using this reselected exterior point $\overline{w}^{\rm ex}$ (shown as the point $EP_2'$ in Fig. \ref{Fig.re-selection}), a corresponding boundary point is updated ($BP_2'$), which will successfully used to obtain a boundary hyperplane. Certainly, if the updated boundary point is still a ``bad" boundary point, repeat solving \eqref{farthest point} to reselect an exterior point and upadte a boundary point.

\subsection{Progressive Geometric Expansion}

Denote $\mathbb{S}^{\rm L, pre}$ as the solved GOSR in previous iteration.
After solving a boundary hyperplane $H: {c'}^{T} w +{d'}\leq 0$ based on an exterior point $w^{ex}$, a new GOSR is updated by $\mathbb{S}^{\rm L,(z)}=\mathbb{S}^{\rm L, pre}\cap S$ (suppose it is the $z$-th iteration of the GOSR solving process, and $S$ is the corresponding inequality constraint of $H$). Then new exterior points ${\hat{w}}^{\rm ex}$ are needed to be found based on the boundary hyperplane $H$ and the previous GOSR $\mathbb{S}^{\rm L, pre}$, which are used to generate other boundary hyperplanes of the target $\mathbb{S}^{\rm L}$. Each  ${\hat{w}}^{\rm ex}$ is sought to satisfy: (1) ${\hat{w}}^{\rm ex}\notin\mathbb{S}^{\rm L}$; (2) ${\hat{w}}^{\rm ex}\in\mathbb{S}^{\rm L,(z)}$; (3) ${\hat{w}}^{\rm ex}$ is an intersection point of $n$ boundary hyperplanes, which include a hyperplane corresponding to $S$ and $n-1$ different hyperplanes corresponding to other inequality constraints selected from the GOSR $\mathbb{S}^{\rm L, pre}$. 
Suppose $\mathbb{S}^{\rm L, pre}:=\{c_i^{T}w+d_{i}\leq 0, \ i=1,2,\dots,K_{\rm pre}\}$, which has $K_{\rm pre}$ corresponding hyperplanes. To generate the mentioned ${\hat{w}}^{\rm ex}$, the exhaustive method is an alternative. Hence, there are $C_{K_{\rm pre}}^{n-1}$ cases of selecting $n-1$ hyperplanes from $K$ hyperplanes. Then combined with $H$, each case will conduct a system of linear equations with $n$ variables and $n$ equations, and it has a unique solution, the solution is a valid intersection. Using the BPS model, if a valid intersection is judged to be an exterior point of $\mathbb{S}^{\rm L}$, this point is a valid ${\hat{w}}^{\rm ex}$. Then all ${\hat{w}}^{\rm ex}$ will be found using this exhaustive method.
However, if there are many boundary hyperplanes in the $\mathbb{S}^{\rm L, pre}$, the cases of the above exhaustive method may be trapped in a combinatorial explosion and its computational time to generate ${\hat{w}}^{\rm ex}$ will take a long time. 

\renewcommand{\algorithmicrequire}{ \textbf{Input:}} 
\renewcommand{\algorithmicensure}{ \textbf{Output:}} 
\begin{algorithm}[tb]
\caption{PHI Algorithm for Solving GOSR}
\label{algorithm-AOSR}
\begin{algorithmic}[1] 
  \REQUIRE ~~\\ 
  parameters of $\mathbb{F}^{\rm L}$ space: $A_{1},A_{2},B_{1},B_{2},F_{1},F_{2}$;\\
  initial data $w^{{\rm ex},(0)}$, $w^{{\rm in},(0)}$, $\mathbb{S}^{{\rm L},(0)}$ and REG dimension $n$;
  \ENSURE ~~\\ 
  GOSR: $\mathbb{S}^{\rm L}$.
  \STATE {Set $\mathbb{S}^{\rm L, pre}=\mathbb{S}^{{\rm L},(0)}$, $W^{\rm ex}=\{w^{{\rm ex},(0)}\}$, $w^{\rm in}=w^{{\rm in},(0)}$, $z=1$;}
  \WHILE {$W^{\rm ex}\neq \varnothing$}
  \STATE {Select a exterior point $w^{\rm ex}$ in the set $W^{\rm ex}$, and the set of the remainder points is $W^{{\rm ex}, \rm re}$};
  \STATE {// \ \textit{OBG method to solve a boundary hyperplane: 5-13}}
    \STATE Solve the boundary point $w^{\rm b}$ by the BPS model in \eqref{BPS model} under $w^{\rm ex}$ and $w^{\rm in}$;
    \FOR{$i=1$ to $n-1$}
    \STATE Solve $i$-th new boundary point $w^{{\rm b_{+}},(i)}$ by the NBPG model model in \eqref{boundary hyperplane Solving model} under $w^{\rm b}$;
    \IF{$\|w_{i}^{{\rm b}_+,(i)}-w_{i}^{\rm b}\| \leq \epsilon$}
    \STATE Reselect an $w^{\rm \overline{ex}}$ by the DEPA model;
    \STATE $w^{\rm ex}\leftarrow w^{\rm \overline{ex}}$ and return to 5;
    \ENDIF
    \ENDFOR
    \STATE {Solve boundary hyperplane $H^{(z)}$ by solving \eqref{HD hyperplane}};
    \STATE {// \ \textit{Ignoring nearly coincident hyperplanes: 15-27}}
    \STATE {Set Flag=0; $K_{\rm pre}=|\mathbb{S}^{\rm L, pre}|$}
    \FOR{$j=1$ to $K_{\rm pre}$}
    \STATE {Compute the cosine $\cos(H^{(z)},H_{S}^{(j)})$ of the angle between $H^{(z)}$ and the $j$-th hyperplane $H_{S}^{(j)}$ of $\mathbb{S}^{\rm L, pre}$;}
    \STATE {// \ \textit{$\phi$ is the maximum tolerated angle}}
    \IF{$\cos(H^{(z)},H_{S}^{(j)}) \geq \cos\phi$} 
    \STATE {Flag=1; \textbf{Break;}}
    \ENDIF
    \ENDFOR
    \IF{Flag=0}
    \STATE Obtain inequality constraints $S^{(z)}$ related to $H^{(z)}$;
    \ELSE
    \STATE {$W^{{\rm ex}}\leftarrow W^{{\rm ex}, \rm re}$ and Return to 3;}
    \ENDIF
    \STATE {$\mathbb{S}^{{\rm L},(z)}\leftarrow \mathbb{S}^{\rm L, pre}\cap S^{(z)}$;}
    \STATE Obtain the set of  ${\hat{w}}^{\rm ex}$ using the PEG model based on $H^{(z)}$ and $\mathbb{S}^{\rm L, pre}$;
    \STATE {//\ \textit{Removing redundant exterior points: 31 }}
    \STATE Update ${\hat{w}}^{\rm ex}$ by deleting the exterior points that are outside $\mathbb{S}^{{\rm L},(z)}$;
    \STATE $W^{{\rm ex}}\leftarrow W^{{\rm ex}, \rm re}\cup{\hat{w}}^{\rm ex}$, $\mathbb{S}^{\rm L, pre}\leftarrow\mathbb{S}^{{\rm L},(z)}$;
    \STATE $z\leftarrow z+1$;
\ENDWHILE
\end{algorithmic}
\end{algorithm}

A progressive geometric expansion (PEG) method is proposed to deal with this issue by progressively expanding the solved geometry, which significantly reduces the computational time of a large scale $\mathbb{S}^{\rm L, pre}$. The proposed PEG method first generates adjacent hyperplanes of hyperplane $H$ by judging only $K_{\rm pre}$ times. If the $k$-th boundary hyperplane of $\mathbb{S}^{\rm L, pre}$ is an adjacent hyperplane of $H$, then there exists at least a point inside $\mathbb{S}^{\rm L, pre}$, which is on the intersecting hyperplane of these two boundary hyperplanes. Therefore, the feasible region of the following constraints \eqref{AH-NEPS model 1} is not empty:
\begin{IEEEeqnarray}{l}
  \label{AH-NEPS model 1}
  \IEEEyesnumber\IEEEyessubnumber*
  c_{i}^{T}w+d_{i}\leq 0, \ i=1,2,\dots,K_{\rm pre}\\
  c_{k}^{T}w+d_{k}= 0,\ c'^{T}w+d^{'}= 0.
\end{IEEEeqnarray}

A linear programming problem can be formed by adding an arbitrary linear objective function (such as simple coordinate summation) to the above constraints \eqref{AH-NEPS model 1}. Then if there is a solution, the $k$-th boundary hyperplane is an adjacent hyperplane of $H$. The advantage of the proposed PEG model is that with the increase of the boundary hyperplane number of $\mathbb{S}^{\rm L, pre}$, the number of adjacent hyperplanes of $H$ will not correspondingly increase dramatically, thereby it will not fall into combinatorial explosion as the exhaustive method.

\subsection{Maximum Tolerated Angle}
In addition, some boundary hyperplanes of GOSR are quite close and almost coincident. Therefore, a special processing method is established to deal with the nearly coincident hyperplanes, which can significantly reduce the computational cost while making little change to the error level. Denote $\phi$ as the maximum tolerated angle of the two different hyperplanes, indicating that two hyperplanes with an angle $\varphi$ (i.e. orange angle in \textcolor{blue}{Fig. \ref{Fig.re-selection}}) smaller than $\phi$ are not distinguished. The cosine of two different hyperplanes is represented by the cosine of the angle between their internal normal vectors,  
\begin{IEEEeqnarray}{l}
  \label{Cos hyperplane}
  \cos(\varphi ) = \cos(H^{(+)},H^{(pre)}) = \frac{\mathbf{n}_{1}\cdot\mathbf{n}_{2}}{\|\mathbf{n}_{1}\|\|\mathbf{n}_{2}\|}
\end{IEEEeqnarray}
where $H^{(+)}$ is the newly generated alternative boundary hyperplane, $H^{(pre)}$ is a previous boundary hyperplane, $\textbf{n}_{H^{(+)}}$ and $\textbf{n}_{H^{(pre)}}$ are the normal vectors of them, and $\Vert n \Vert$ means the modulus length of the normal vector.

\subsection{Complete Process of PHI Algorithm}
Firstly, the following preparations are required: 

\subsubsection{Initial Box Region of GOSR}
The initial box region of GOSR are set to $\mathbb{S}^{\rm L, (0)}:=\{w|\ 0\leq w_{i}\leq w_{i}^{\max}, \ i\in \mathcal{R}\}$, where $w_{i,\max}$ is the upper limitation of REG output at node $i$. 
The flexible $w_{i}^{\max}$ can be set to the current capacity of REG at node $i$, or the capacity for future construction, or a larger capacity preset to test capacity boundary. 

\subsubsection{Initial Interior Point}
The initial interior point can be set to the coordinate origin, given by $w^{\rm in, (0)}=\{w_{i}^{\rm in, (0)}=0, \ i\in\mathcal{R}\}$, which means that the power system is secure when there is zero output of REG. 

\subsubsection{Initial Exterior Point}
The initial exterior point is set to the farthest point in the initial box region from the initial interior point, i.e. each REG output reaches its upper limitations, given by $w^{\rm ex,(0)}=\{w_{i}^{\max}, \ i\in\mathcal{R}\}$.

The PHI algorithm progressively obtains all inequality expressions of the GOSR by the iterative process as ``exterior point $\rightarrow$ boundary hyperplane $\rightarrow$ new exterior point $\rightarrow \dots$", where $\epsilon$ is the error for judging whether two points are the same point that can be set to $10^{-6}$. \textbf{Algorithm \ref{algorithm-AOSR}} gives the complete process of the PHI algorithm in pseudocode.

\section{Case Study}
\subsection{Test Networks}
The proposed PHG method is tested in two power systems: 1) the IEEE 30-bus transmission network \cite{ieee30bus} with seven generators; 2) the IEEE 118-bus transmission network \cite{ieee118}: which hosts 19 generators, and the cost coefficients of generators are presented in \cite{blumsack2006network}. In the proposed PHG method, the obtained GOSR $\mathbb{S}^{\rm L}$ should be the error-free and accurate projection of the linearized feasible region $\mathbb{F}^{\rm L}$ if the maximum tolerated angle $\phi=0$. Therefore, the coincidence of $\mathbb{S}^{\rm L}$ and $\mathbb{F}^{\rm L}$ is the indicator of the accuracy performance. In addition, this case study also focuses on the accuracy of the $\mathbb{S}^{\rm L}$ compared to the feasible region of the AC power system ($\mathbb{F}^{\rm AC}$). The coincidence is quantified by the Monte Carlo simulation (MCS) method, which represents the real operational security region of the power system. 
The benchmark cases of MCS based on ACOPF take ${10}^{N_{\rm d}+1}$ MCS points respectively, where $N_{\rm d}$ denotes the dimensions of REG. The benchmark cases of the enumeration method refer to \cite{huynh_practical_1992, wan_maximum_2018}. The benchmark cases of the Disp-Reg method that iteratively seeks MILP optimization refer to \cite{wei_dispatchable_2015,wang_sequential_2021,liu_data-driven_2022}. The chosen nodes of IEEE 118-bus are [2,5,8,111] and that of IEEE 30-bus are [5,7,9,21]. All the cases are implemented in Matlab R2021b on a laptop with AMD-Ryzen7-5800H CPU and 16GB of RAM.

\begin{figure}[tb]
  \centering
  \subfigure[IEEE 30-bus]{
  \centering
  \label{Fig.2D_TN33} 
  \includegraphics[width=3.2in]{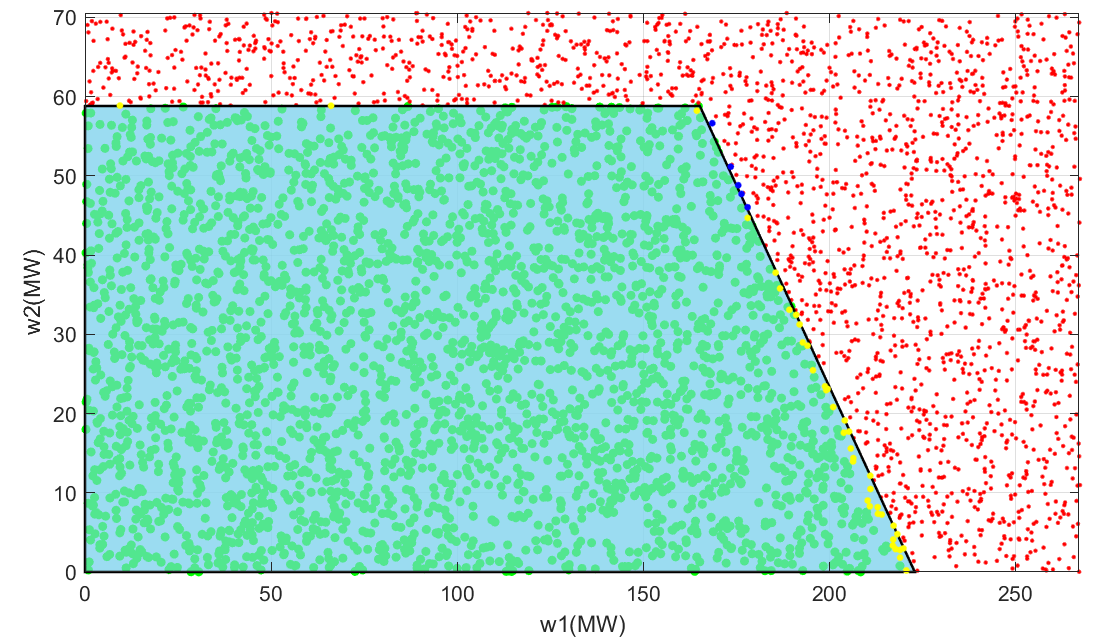}}
  \subfigure[IEEE 118-bus]{
  \centering
  \label{Fig.2D_TN118} 
  \includegraphics[width=3.2in]{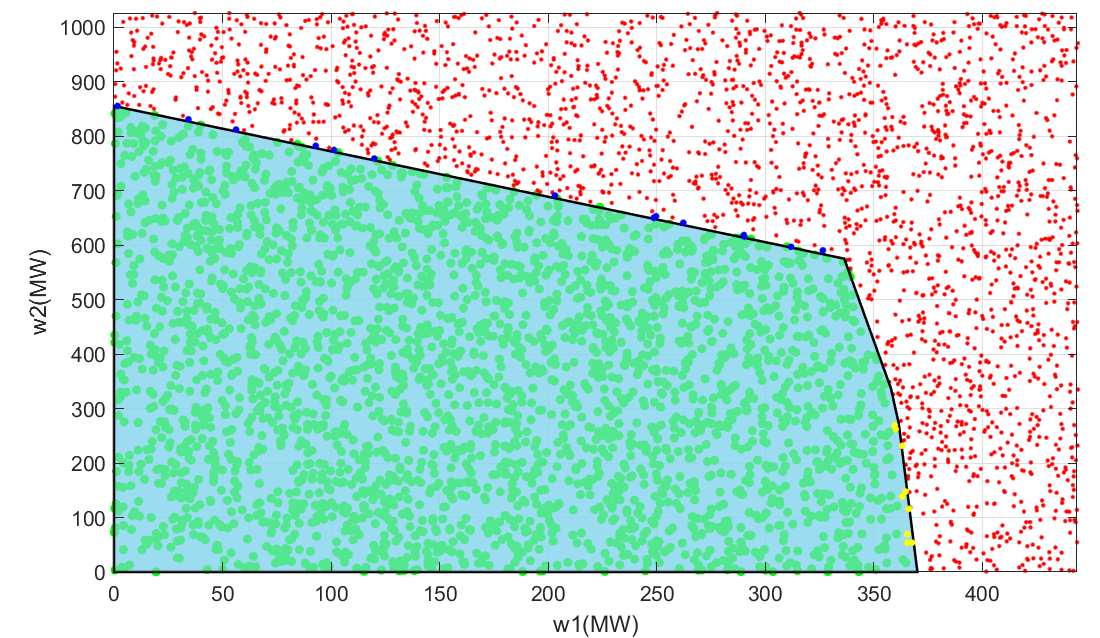}}
  \caption{Visualization results of PHG method and verified by MCS: 2D}
  \label{Fig.2D_Accuracy} 
\end{figure}

\begin{figure}[tb]
  \centering
  \subfigure[IEEE 30-bus]{
  \centering
  \label{Fig.3D_TN33} 
  \includegraphics[width=3.33in]{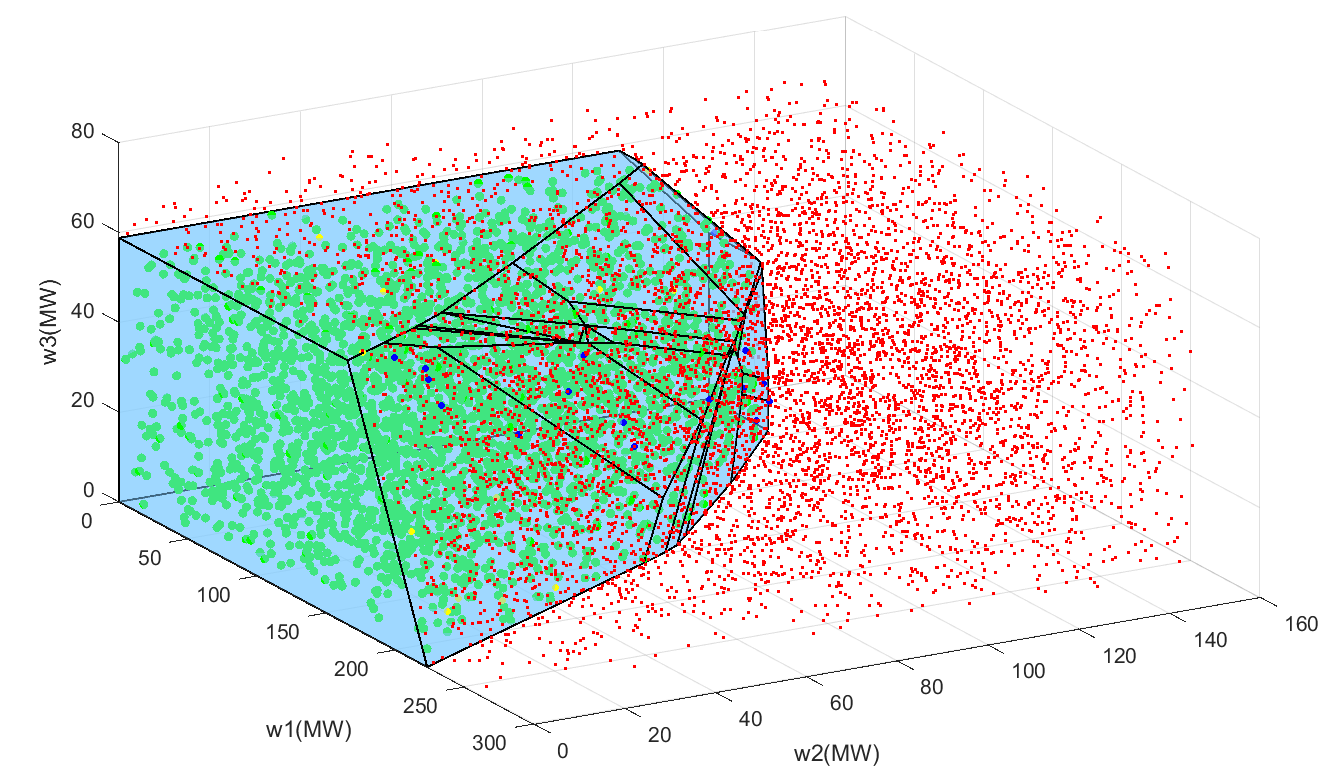}}
  \subfigure[IEEE 118-bus]{
  \centering
  \label{Fig.3D_TN118} 
  \includegraphics[width=3.33in]{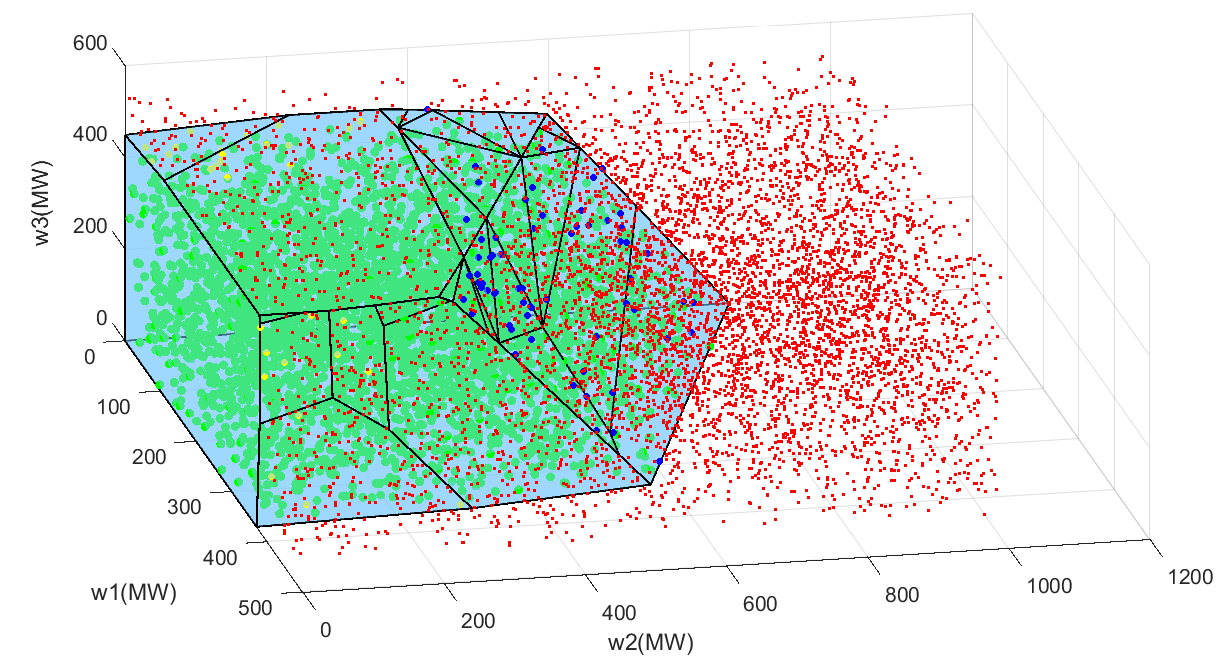}}
  \caption{Visualization results of PHG method and verified by MCS: 3D}
  \label{Fig.3D_Accuracy} 
\end{figure}

\subsection{Visualization of GOSR and Accuracy Metric}
\textcolor{blue}{Fig. \ref{Fig.2D_Accuracy}} and \textcolor{blue}{Fig. \ref{Fig.3D_Accuracy}} visualize the real experimental results of the GOSR $\mathbb{S}^{\rm L}$ solved by the proposed PHG method, where the numbers of REG nodes in the IEEE 30-bus and IEEE 118-bus systems are two (2D) and three (3D) respectively. Denote $\mathbb{F}^{\Phi}$ as the benchmark region, i.e. feasible region $\mathbb{F}^{\rm AC}$ or linearized feasible region $\mathbb{F}^{\rm L}$. The color of the MCS samples in the figures means that: 

1) Green: the sample point is inside $\mathbb{S}^{\rm L}$ and $\mathbb{F}^{\Phi}$, thus forming the set $\{w |w\in \mathbb{S}^{\rm L}, \exists x \rightarrow (w, x) \in \mathbb{F}^{\Phi}\}$;

2) Blue: the sample point is outside $\mathbb{S}^{\rm L}$ and inside $\mathbb{F}^{\Phi}$, thus forming the set $\{w |w \notin \mathbb{S}^{\rm L}$,  $\exists x \rightarrow (w, x) \in \mathbb{F}^{\Phi}\}$;

3) Yellow: the sample point is inside $\mathbb{S}^{\rm L}$ and outside $\mathbb{F}^{\Phi}$, thus forming the set $\{w |w\in \mathbb{S}^{\rm L}$,  $\nexists x \rightarrow (w, x) \in \mathbb{F}^{\Phi}\}$;

4) Red: the sample point is outside $\mathbb{S}^{\rm L}$ and $\mathbb{F}^{\Phi}$, thus forming the set $\{w |w\notin \mathbb{S}^{\rm L}$,  $\nexists x \rightarrow (w, x) \in \mathbb{F}^{\Phi}\}$.

The error can be formulated by, 
\begin{IEEEeqnarray}{l}
  \label{error}
  E_{\rm r}=(1-N_{\rm SA}/N_{\rm SR})100\%
\end{IEEEeqnarray}
where $N_{\rm SR}$ is the number of samples inside $\mathbb{S}^{\rm L}$ and $N_{\rm SA}$ is the number of samples which behave the same in two regions.

Based on the different benchmark regions, the geometry error and the approximation error are defined.
The geometry error demonstrates the accuracy of the tested method compared to linearized feasible region $\mathbb{F}^{\rm L}$, which represents how well the final constructed geometry matches the real projection of $\mathbb{F}^{\rm L}$ onto the space of REG.
The approximation error evaluates the approximation accuracy of the tested method compared to feasible region $\mathbb{F}^{\rm AC}$, which represents how well the final constructed geometry approximates the real projection of $\mathbb{F}^{\rm AC}$ onto the space of REG (implicit and nonlinear IOSR).

\begin{figure}[tb]
  \hspace*{-0.3cm}
  \label{Fig.Sen_Error_3D} 
  \includegraphics[width=3.8in]{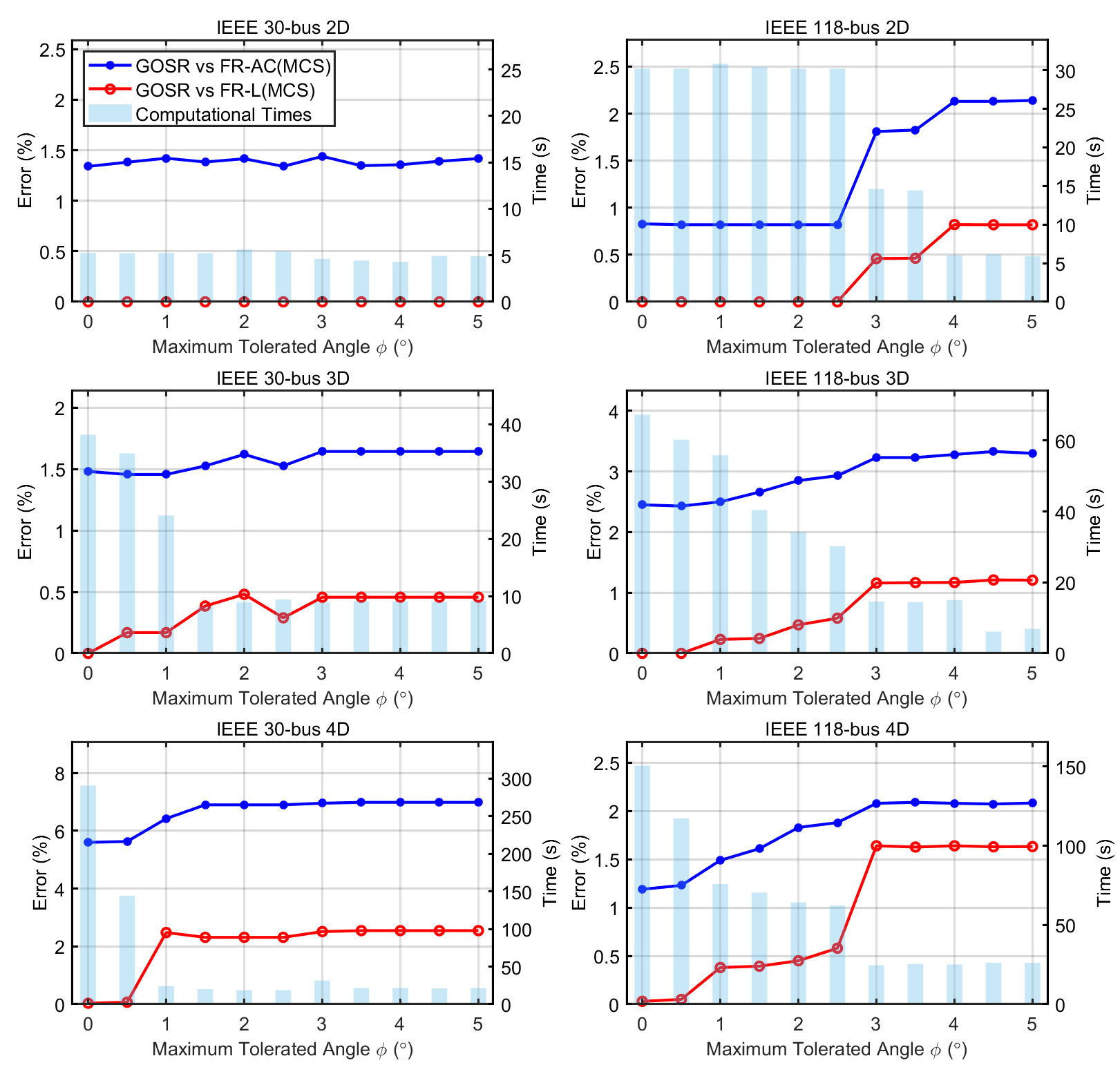}
  \caption{Errors of PHG method under different values of $\phi$.}
  \label{Fig.Sen_Error} 
\end{figure}

\subsection{Sensitivity Analysis of Maximum Tolerated Angle}
\textcolor{blue}{Fig. \ref{Fig.Sen_Error} } shows the variation of calculational time and errors under different values of $\phi$ of the proposed PHG method. Generally, the geometry and approximation errors of the PHG method exhibit a fluctuating positive trend with the increase of $\phi$, then fluctuating at a stable value. Under this trend, the IEEE 30-bus 3D and 4D cases have abnormal changes. Specifically, the error in IEEE 30-bus 4D jumps to a high value at a certain angle $\underline{\phi}$ between 0.5° and 1° in and fluctuates slightly to a threshold angle $\overline{\phi}$ around 3°, beyond which it tends to stabilize at a specific value meaning further $\phi$ increase does not result in significant accuracy changes. Conversely, the error jumping at $\underline{\phi}$ is not relatively significant in IEEE 30-bus 3D, while the error holding beyond $\overline{\phi}$ is relatively significant. The phenomenon is more pronounced in high-dimensional spaces where the errors of the linearized ACOPF model and the PHG method exhibit exponential growth compared to the lower-dimensional space of REGs. 

From the geometric foundation of the PHG method, it indicates that the angle between adjacent boundary hyperplanes should be concentrated between a lower bound $\underline{\phi}$ and an upper bound $\overline{\phi}$. When $\phi$ is set within this range, the error shows high sensitivity, whereas the error can be considered insensitive to $\phi$ outside this range. The high $\phi$ values are not considered here, which will cause the model to be too inaccurate. 
The results show that 0° to 3° of $\phi$ is the more sensitive range, which will be further discussed in the subsequent accuracy analysis. Obviously, in the $\phi$-sensitive range, the calculation time will also be reduced more significantly. In practice, users can flexibly balance accuracy and computational efficiency. 

\begin{table}[tb]
  \renewcommand{\arraystretch}{1.3}
  \caption{ Errors of The PHG Method and Benchmarks}
  \centering
  \setlength{\tabcolsep}{1.2pt} 
  \begin{minipage}{\linewidth}
  \begin{tabular}{c|cccccccc}
  \hline
  \multirow{2}{*}{} & \multirow{2}{*}{Network} & \multirow{2}{*}{$N_{\rm d}$} & \multicolumn{4}{c}{PHG Method} & \multirow{2}{*}{Disp-Reg} & \multirow{2}{*}{Enum.} \\ 
  & & & $\phi=0°$ & 1° & 2° & 3° &  &  \\ 
  \hline
  \multirow{6}{*}{\shortstack{Geometry \\ Errors \\ ($\mathbb{S}^{\rm L}$ vs $\mathbb{F}^{\rm L}$)}} 
  & \multirow{3}{*}{\shortstack{IEEE \\30-bus}}
  & 2D & 0\% & 0\%    & 0\%    & 0\%    & 1.73\% & 0\% \\
  & & 3D & 0\% & 0.21\% & 0.48\% & 0.46\% & 0.61\% & 0\% \\
  & & 4D & 0\% & 2.48\% & 2.31\% & 2.51\% & 0.47\% & 0\% \\
  \cline{2-9}
  & \multirow{3}{*}{\shortstack{IEEE \\118-bus}}
  & 2D & 0\%    & 0\%    & 0\%    & 0.46\% & 0.03\% & 0\% \\
  & & 3D & 0\%    & 0.23\% & 0.47\% & 1.16\% & 0.13\% & 0\% \\
  & & 4D & 0\% & 0.38\% & 0.45\% & 1.64\% & 0.05\% & 0\% \\
  \cline{1-9}
  \multirow{6}{*}{\shortstack{Approximation \\ Errors \\ ($\mathbb{S}^{\rm L}$ vs $\mathbb{F}^{\rm AC}$)}} 
  & \multirow{3}{*}{\shortstack{IEEE \\30-bus}}
  & 2D & \textbf{1.34\%} & 1.34\% & 1.34\% & 1.34\% &\textbf{4.09\%} &\textbf{2.34\%} \\
  & & 3D & \textbf{1.48\%} & 1.46\% & 1.62\% & 1.64\% &\textbf{8.89\%} &\textbf{8.22\%} \\
  & & 4D & \textbf{5.60\%} & 6.42\% & 6.90\% & 6.96\% &\textbf{11.40\%} &\textbf{10.86\%} \\
  \cline{2-9}
  & \multirow{3}{*}{\shortstack{IEEE \\118-bus}}
  & 2D & \textbf{0.82\%} & 0.82\% & 0.82\% & 1.81\% &\textbf{2.36\%} &\textbf{2.29\%} \\
  & & 3D & \textbf{2.45\%} & 2.50\% & 2.85\% & 3.23\% &\textbf{3.45\%} &\textbf{3.30\%} \\
  & & 4D & \textbf{1.19\%} & 1.49\% & 1.83\% & 2.08\% &\textbf{3.57\%} &\textbf{3.62\%} \\
  \hline
  \end{tabular}
  \end{minipage}
  \label{Errors of proposed PHG method}
\end{table}

\subsection{Accuracy Analysis of GOSR}

\subsubsection{Geometry Errors}
\textcolor{blue}{Table \ref{Errors of proposed PHG method}} details the geometry errors and approximation errors of the PHG method in various $\phi$ with benchmarks, where the enumeration method (abbreviated as Enum.) is geometry error-free in principle.
The results show that the geometry errors of the PHG method remain nearly 0\% when $\phi=0^\circ$, compared to the Disp-Reg method that exhibits geometry errors ranging from 0.05\% to 1.73\%. The small geometry errors are caused by the relaxation of MILP problems and inherent errors in the bisection step of the benchmark method. Theoretically, when $\phi = 0$, the proposed PHG method can precisely obtain all boundary hyperplanes, except for cases involving extremely close hyperplanes. The accuracy of the PHG method can be explained by the following two reasons. It solves the exact global optimal solution of linear programming in OBG. And finite boundary hyperplanes of geometry can be completely obtained by the PHI algorithm. Besides, this accuracy may diminish due to discarding some close hyperplanes as $\phi$ increases to reduce the computational burden. This is because the lack of some boundaries will cause some infeasible points to be included wrong. Evidently, the results validate the precision of the PHG method. 

\subsubsection{Approximation Errors}
The results of approximation errors indicate that the proposed PHG method consistently achieves a significant reduction in approximation error across varying network scales. For the 30-bus case, the PHG method outperforms the benchmark by nearly halving or more compared to the benchmark, with a maximum reduction of 7.41\% from 8.89\% of the benchmark to 1.48\% in the 3D case. A similar trend is observed in the 118-bus case, showcasing the scalability and effectiveness of the PHG method across both small-scale and large-scale systems.  
In addition, the results show that the superiority of approximation errors is maintained as the dimensionality increases, which verify the PHG method is applicable to projection spaces of multiple dimensions.
Moreover, the PHG method exhibits remarkable stability under small incremental variations in $\phi$, maintaining superior accuracy over the benchmark across all test cases. The maximum approximation error increase value of 1.36\% from 5.60\% at $\phi=0^\circ$ to 6.96\% at $\phi=3^\circ$ in 30-bus 4D case, which is still far better than the Disp-Reg with 11.4\% and enumeration method with 10.86\%. 
This stability, combined with its consistent and strong approximation accuracy to IOSR, demonstrates the reliability of the PHG method.

\begin{table}[tb]
  \renewcommand{\arraystretch}{1.3}
  \caption{Computational Costs (s)}
  \centering
  \setlength{\tabcolsep}{3.3pt} 
  \begin{center}
    \begin{tabular}{cccccc}
      \hline
      \multirow{2}{*}{$N_{\rm d}$} & \multirow{2}{*}{Methods} & \multicolumn{2}{c}{IEEE 30-bus} & \multicolumn{2}{c}{IEEE 118-bus} \\
      \cline{3-6}
      & & $( N_{\rm all}, N_{\rm new})$ & Time (s) & $( N_{\rm all}, N_{\rm new})$ & Time (s) \\
      \hline
      \multirow{4}{*}{2D} 
      & PHG & (6,4) & 7.8720  & (15,13) & \textbf{30.2038} \\
      & Disp-Reg & (5,3) & 15.2509 & (13,11) & 60.2811 \\
      & Enumeration & (250,4) & \textbf{1.47738} & (1054,13) & 32.7501 \\
      & MCS & N/A & 361.3216 & N/A & 433.019 \\
      \hline
      \multirow{4}{*}{3D} 
      & PHG & (15,12) & 34.0002 & (21,18) & \textbf{67.2590} \\
      & Disp-Reg & (9,6) & \textbf{24.3279} & (16,13) & 106.4699 \\
      & Enumeration & (250,12) & 178.1811 & (1054,18) & 211.0036 \\
      & MCS & N/A & 3513.9303 & N/A & $>$1h \\
      \hline
      \multirow{4}{*}{4D} 
      & PHG & (79,75) & \textbf{377.6875} & (44,40) & \textbf{150.5056} \\
      & Disp-Reg & (101,97) & 1154.9752 & (51,47) & 693.0945 \\
      & Enumeration & (250,75) & 1632.5801 & (1054,40) & $>$1h \\
      & MCS & N/A & $>$1h & N/A & $>$1h \\
      \hline
    \end{tabular}
  \end{center}
  \label{Computational Costs}
\end{table}

\subsection{Computational Efficiency Analysis}

The computational costs of the PHG method compared to benchmark methods are given in \textcolor{blue}{Table \ref{Computational Costs}}, where $( N_{\rm all}, N_{\rm new})$ are the number of all inequalities and new valid hyperplanes (except initial boundary) respectively of GOSR $\mathbb{S}^{\rm L}$. 
The results show that the enumeration and Disp-Reg methods exhibit relatively better computational efficiency in lower-dimensional REG with fewer hyperplanes to be solved, such as in IEEE 30-bus 2D and 3D cases. And the MCS method always has the maximum computational burden. However, when more hyperplanes should be solved, the proposed PHG method demonstrates higher computational efficiency. Particularly in the bus-30 4D case, the PHG method reduces computational time from 1632.58s to 377.69s by 76.87\% compared to the enumeration method, and from 693.09s to 150.51s by 78.29\% compared to the Disp-Reg method in the 118-bus 4D case. 

The higher efficiency of the proposed PHG method can be attributed to the following reasons. The computational time is positively correlated to the number of boundary hyperplanes, which is much larger in the 4D cases than in lower-dimensional cases, just like 75 new 4D hyperplanes vs. 4  new 2D hyperplanes in the 30-bus cases.  Traditional methods are more effective when the system is small with few constraints or low projected dimensionality. When traditional methods solve a large number of hyperplanes, the number of constraint combinations is quite large after solving a boundary hyperplane each time. By contrast, the proposed PEG method solves the exterior points where the new boundary hyperplane intersects with the adjacent hyperplanes. It significantly decreases the number of combinations and reduces computation time, which is more suitable for the high-dimensional REG space with large numbers of boundary hyperplanes.

\section{Conclusion}
The geometric expression of the operational security region of REG is convenient for assessing the safe operation risks of power systems. This paper proposes a novel PHG method to obtain it as a geometry constructed by boundary hyperplanes progressively. 
To begin with, an OBG method is proposed to solve a boundary hyperplane of the target polytope models as finite linear programming problems based on intersecting and orthogonal geometric relationships. 
Furthermore, a PHI algorithm is built to effectively obtain all boundary hyperplanes. Besides, the proposed DEPA method ensures iterative continuity. And the PEG method prevents combinatorial explosion of hyperplanes, thereby guaranteeing computational efficiency.
Moreover, comprehensive case studies demonstrate geometry and approximation accuracy, flexible trade-offs between accuracy and efficiency, as well as adaptability to computational cost in both small- and large-scale systems of the proposed method.
In general,  the proposed PHG method establishes an effective framework for the operational security region of renewable energy generation in power systems.

\appendices
\section{Proof of Proposition 1}
\label{Proof of Proposition 1}
\textit{Proposition 1}: Suppose the given point $w^{\rm gp}$ is an exterior point of the GOSR $\mathbb{S}^{\rm L}$, and there exists a necessary and sufficient condition in \eqref{relation of V-AC and S-AC}, then the point $w^{\rm b}$ solved by \eqref{BPS model} is a point on a boundary hyperplane of $\mathbb{S}^{\rm L}$.

\textit{Proof}: Denote $(\lambda^{*},x^{*})$ is the global optimal solution solved by \eqref{BPS model}. Because the given point $w^{\rm gp}$ is an exterior point $w^{\rm ex}$ of $\mathbb{S}^{\rm L}$, there exists $0<\lambda^{*}<1$ and $w^{\rm b}=w^{\rm in}+(w^{\rm ex}-w^{\rm in})\lambda^{*}$. It is seen from \eqref{BPS constraints} that $(w^{\rm b},x)\in\mathbb{F}^{\rm L}$, thereby $w^{\rm b}$ also satisfies $w^{\rm b}\in\mathbb{S}^{\rm L}$.

Assume that the boundary point $w^{\rm b}$ is not a point on any boundary hyperplane of $\mathbb{S}^{\rm L}$. Since $w^{\rm b}$ satisfies $w^{\rm b}\in\mathbb{S}^{\rm L}$, the point $w^{\rm b}$ is a strict interior point. Then there exists $\delta>0$ and $\delta+\lambda^{*}<1$ makes any point in the neighbourhood $U(w^{\rm b},\delta)=\{w|\,\|w-w^{\rm b}\|<\delta\}$ satisfing $w\in\mathbb{S}^{\rm L}$.
\begin{align}
  \label{Neighbourhood interior}
  w\in U(w^{\rm b},\delta) \Rightarrow w\in\mathbb{S}^{\rm L}
\end{align}

Define $\lambda^{\delta}=\lambda^{*}+\frac{0.5\delta}{\|w^{\rm gp}-w^{\rm in}\|}$. It can be seen clearly that $\lambda^{*}<\lambda^{\delta}<1$. The new point $w^{\delta}=w^{\rm in}+(w^{\rm ex}-w^{\rm in})\lambda^{\delta}$ satisfies $\|w^{\delta}-w^{\rm b}\|=0.5\delta<\delta$, i.e. $w^{\delta}\in U(w^{\rm b},\delta)$. It can be indicated from \eqref{Neighbourhood interior} that the point $w^{\delta}\in\mathbb{S}^{\rm L}$, thereby there exist $w^{\delta}$ have $(w^{\delta},x^{\delta})\in\mathbb{F}^{\rm L}$. Then $w^{\delta}$ is also a feasible point that satisfies the constraints in \eqref{BPS model}. Then from $\lambda^{*}<\lambda^{\delta}$, it is shown that the solution $\lambda^{\delta}$ is a better solution than $\lambda^{*}$, which is contradicted with the situation that $\lambda^{*}$ is a global optimum. Obviously, the assumption that the nearest interior point $w^{\rm b}$ is not a point on any boundary hyperplane is contradictory. Therefore, the boundary point $w^{\rm b}$ is a point on a boundary hyperplane of $\mathbb{S}^{\rm L}$. This ends the proof of \textit{Proposition 1}.

\section{Proof of Theorem 1}
\label{Proof of Theorem 1}
\textit{Theorem 1}: The necessary and sufficient condition for using $n$ points ($\alpha_{n},\ i=1,2,\dots,n$) to uniquely determine a hyperplane in the $n$-dimensional linear space is that $n-1$ basis vectors $\alpha_{2}-\alpha_{1},\alpha_{3}-\alpha_{1},\dots,\alpha_{n}-\alpha_{1}$ are linearly independent ($n-1$ orthogonal basis).

\textit{Proof}: A hyperplane can be formulated as $c^T \alpha+d=0$, where the dimensions of $c$ and $d$ are $n$ and 1, respectively. Then the hyperplane passing through these $n$ points satisfies $c^T \alpha_{i}+d=0,\ i=1,2,\dots,n$, written in matrix form as: 
\begin{align}
  \label{linear eqaution to solve hyperplane}
  M  
  \left[
    \begin{array}{c}
      c\\
      d
    \end{array}
  \right]
  =
  \left[
  \begin{array}{cc}
    (\alpha_{1})^{T} &1 \\
    \vdots & \vdots\\
    (\alpha_{n})^{T} &1
  \end{array}
  \right]
  \left[
  \begin{array}{l}
    c\\
    d
  \end{array}
  \right]
  =0
\end{align}
where $M$ is an $n$ row and $n+1$ column matrix. 

The  homogeneous system of linear equations \eqref{linear eqaution to solve hyperplane} is used to determine the hyperplane $c^T \alpha+d=0$ by solving variables $c$ and $d$, which are the non-zero solutions of \eqref{linear eqaution to solve hyperplane}.

Introduce a matrix $M'$ defined by \eqref{M' equation}. The rank of $M'$ is equal to that of $M$, i.e. $R(M')=R(M)$.
\begin{align}
  \label{M' equation}
  M'\hspace{-0.3em}=\hspace{-0.3em}
  \left[\hspace{-0.3em}
  \begin{array}{ccccc}
    1 &0 &0 & \dots& 0\\
    -1 &1 &0 & \dots& 0\\
    \vdots &\vdots &\vdots &\vdots& \vdots\\
    -1 &0 &0 & \dots& 1\\
  \end{array}\hspace{-0.3em}
  \right]
  \hspace{-0.3em}M\hspace{-0.3em}
  =\hspace{-0.3em}
  \left[
    \hspace{-0.3em}
  \begin{array}{cc}
    (\alpha_{1})^{T} &1 \\
    (\alpha_{2}-\alpha_{1})^{T} &0 \\
    \vdots & \vdots\\
    (\alpha_{n}-\alpha_{1})^{T} &0
  \end{array}
  \hspace{-0.3em}
  \right]
\end{align}

\subsubsection{Proof of Sufficient Condition} 
If the vectors $\alpha_{2}-\alpha_{1},\alpha_{3}-\alpha_{1},\dots,\alpha_{n}-\alpha_{1}$ are linearly independent, then there exists $R(M)=R(M')=n$. Then the non-zero solution set of \eqref{linear eqaution to solve hyperplane} can be expressed as $[c,d]=r[c_1, d_1],\ \forall r\neq0$, where $[c_1, d_1]$ is the fundamental set of solutions. Obviously, the hyperplance expression $r(c_1^T w +b_1)=0, \forall r\neq0$ refers to a unique hyperplane $c_1^T w +b_1=0$. Therefore, the hyperplance are uniquely determined using these $n$ points if the vectors $\alpha_{2}-\alpha_{1},\alpha_{3}-\alpha_{1},\dots,\alpha_{n}-\alpha_{1}$ are linearly independent.

\subsubsection{Proof of Necessary Condition}
If the hyperplane solved by using $n$ points are uniquely, the non-zero solution set of \eqref{linear eqaution to solve hyperplane} only can be expressed as $[c,d]=r[c_1, d_1],\ \forall r\neq0$, thereby the rank of $M$ must be $n$. Then $R(M')=R(M)=n$. It is indicated from \eqref{M' equation} that the vectors $\alpha_{2}-\alpha_{1},\alpha_{3}-\alpha_{1},\dots,\alpha_{n}-\alpha_{1}$ must be linearly independent. Therefore the vectors $\alpha_{2}-\alpha_{1},\alpha_{3}-\alpha_{1},\dots,\alpha_{n}-\alpha_{1}$ are linearly independent if the hyperplance are uniquely solved using these $n$ points.
This ends the proof of \textit{Theorem 1}.

\section{Proof of Proposition 2}
\label{Proof of Proposition 2}
\textit{Proposition 2}: Suppose that $w^{\rm b}$ is a boundary point of the GOSR $\mathbb{S}^{\rm L}$ and the necessary and sufficient condition in \eqref{relation of V-AC and S-AC} holds. If there exists a point $w^{{\rm b}_+}$ satisfying $(w^{{\rm b}_+}, x^{+})\in\mathbb{F}^{\rm L}$, $(w^{{\rm b}_-}, x^{-})\in\mathbb{F}^{\rm L}$, and $w^{{\rm b}_-}=2 w^{\rm b}-w^{{\rm b}_+}$, where $w^{{\rm b}_-}, x^{-}$ is the auxiliary point, then the new point $w^{{\rm b}_+}$ is also on the hyperplane of $\mathbb{S}^{\rm L}$ passing through the boundary point $w^{\rm b}$.

\textit{Proof}: Since $(w^{{\rm b}_+}, x^{+})\in\mathbb{F}^{\rm L}$ and $(w^{{\rm b}_-}, x^{-})\in\mathbb{F}^{\rm L}$ are satisfied, it can be indicated from the necessary and sufficient condition in \eqref{relation of V-AC and S-AC} that there also have $w^{{\rm b}_+}\in\mathbb{S}^{\rm L},\quad w^{{\rm b}_-}\in\mathbb{S}^{\rm L}$.

Denote $c^{T} w+d=0$ as the hyperplance of $\mathbb{S}^{\rm L}$ passing through $w^{\rm b}$, and $c^{T} w+d\leq0$ is the corresponding inequality constraint of $\mathbb{S}^{\rm L}$. It is indicated that the inequalities $c^{T} w^{{\rm b}_+}+d\leq 0$ and $c^{T} w^{{\rm b}_-}+d\leq 0$ are also satisfied. Since the hyperplane $c^{T} x+d=0$ passing through $w^{\rm b}$, we have $c^{T} w^{\rm b}+d=0$. Then the following inequality is derived.
\begin{align}
  c^{T} w^{{\rm b}_-}+d=&c^{T} (2w^{\rm b}-w^{{\rm b}_+})+d\notag\\ 
  =&2 (c^{T} w^{\rm b}+d) -c^{T} w^{{\rm b}_+}+ d\notag\\ 
  =&-c^{T} w^{{\rm b}_+}+ d \leq 0
\end{align}

It is indicated from $c^{T} w^{{\rm b}_+}+d\leq 0$ and $-c^{T} w^{{\rm b}_+}+ d \leq 0$ that the equation $c^{T} w^{{\rm b}_+}- d = 0$ holds. Therefore, the point $w^{{\rm b}_+}$ is also on the hyperplane of $\mathbb{S}^{\rm L}$ passing through $w^{\rm b}$. This ends the proof of \textit{Proposition 2}.

\section{Proof of Lemma 1}
\label{Proof of Lemma 1}
\textit{Lemma 1}: Suppose points $w^{{\rm b}_+,(i)},\ i=1,\dots,n-1$ in an $n$-dimensional linear space satisfy:

\textit{L1}: $w^{{\rm b}_+,(i)}$ is the $i$-th new boundary point solved by \eqref{boundary hyperplane Solving model};

\textit{L2}: $\|w^{{\rm b}_+,(i)}-w^{\rm b}\| > 0,\ \forall i=1,\dots,n-1$.

Then the hyperplane is uniquely solved by \eqref{HD hyperplane}, which is also the boundary hyperplane of $\mathbb{S}^{\rm L}$ passing through the boundary point $w^{\rm b}$.

\textit{Proof}: Since each point is solved by \eqref{boundary hyperplane Solving model}, it is found from \textit{Proposition 2} that the constraints in \eqref{Constraint1:points on hyperplane}-\eqref{Constraint3:points on hyperplane} ensure that the points $w^{{\rm b}_+,(i)},\ i=1,\dots,n-1$ are all on the boundary hyperplane of $\mathbb{S}^{\rm L}$ passing through $w^{\rm b}$. 

It is indicated from condition \textit{L2}$:\|w^{{\rm b}_+,(i)}-w^{\rm b}\|\geq 0,\ \forall i=1,\dots,n-1$, which means they are all non-zero vectors. The orthogonal constraints in \eqref{orthogonal constraints} guarantees that $w^{{\rm b}_+, (i)}-w^{\rm b}$ are orthogonal vectors, which means these vectors are linearly independent. Therefore, the necessary and sufficient condition in \textit{Theorem 1} is satisfied by points $w^{{\rm b}_+,(i)}$, which uniquely determine a hyperplane passing through these points.

It is found that the hyperplane passing through these points are uniquely determined, and all points are on the boundary hyperplane of $\mathbb{S}^{\rm L}$ passing through $w^{\rm b}$. Therefore, the uniquely solved hyperplane is also the boundary hyperplane of $\mathbb{S}^{\rm L}$ passing through $w^{\rm b}$. This ends the proof of \textit{Lemma 1}.

\bibliographystyle{IEEEtran}
\bibliography{Reference}
\end{document}